\documentclass[aip, jmp, amsmath, amssymb, reprint, author-year]{revtex4-1}
\usepackage{graphicx}
\usepackage{bm}
\usepackage[mathlines]{lineno}
\usepackage[colorlinks=true, linkcolor=blue,citecolor=blue, urlcolor=blue]{hyperref}

\newcommand{\be}{\begin{equation}}
\newcommand{\ee}{\end{equation}}
\newcommand{\beq}{\begin{eqnarray}}
\newcommand{\eeq}{\end{eqnarray}}
\def\ba{\begin{array}}
\def\ea{\end{array}}
\def\bd{\begin{displaymath}}
\def\ed{\end{displaymath}}

\def\bc{\begin{center}}
\def\ec{\end{center}}

\begin{document}


\title[]{Electromagnetic ion cyclotron instability stimulated by the suprathermal ions in space plasmas: A quasi-linear approach}

\author{S.M. Shaaban}
 \email{s.m.shaaban88@gmail.com}
\affiliation{Institute of Experimental and Applied Physics, University of Kiel, Leibnizstrasse 11, D-24118 Kiel, Germany}
\affiliation{Theoretical Physics Research Group, Physics Dept., Faculty of Science, Mansoura University, 35516 Mansoura, Egypt}
\affiliation{Centre for Mathematical Plasma-Astrophysics, KU Leuven, Celestijnenlaan 200B, 3001 Leuven, Belgium}

\author{M. Lazar}%
 \affiliation{Centre for Mathematical Plasma-Astrophysics, KU Leuven, Celestijnenlaan 200B, 3001
Leuven, Belgium}%
 \affiliation{Institut f\"ur Theoretische Physik, Lehrstuhl IV: Weltraum- und Astrophysik, Ruhr-Universit\"at Bochum, D-44780 Bochum, Germany}%
 
\author{R. Schlickeiser}%
 \affiliation{Institut f\"ur Theoretische Physik, Lehrstuhl IV: Weltraum- und Astrophysik, Ruhr-Universit\"at Bochum, D-44780 Bochum, Germany}
 \affiliation{Institut f\"ur Theoretische Physik and Astrophysik, Christian-Albrechts-Universit\"at Kiel, Leibnitzstr. 15, D-24118 Kiel, Germany}

\date{\today}

\begin{abstract}
In collision-poor space plasmas protons with an excess of kinetic energy or temperature in direction perpendicular to background magnetic field can excite the electromagnetic ion cyclotron (EMIC) instability. This instability is expected to be highly sensitive to suprathermal protons, which enhance the high-energy tails of the observed velocity distributions and are well reproduced by the (bi-)Kappa distribution functions. 
In this paper we present the results of a {refined} quasilinear (QL) approach, able to describe the effects of suprathermal protons on the extended temporal evolution of EMIC instability. It is thus shown that suprathermals have a systematic stimulating effect on the EMIC instability, enhancing not only the growth rates and the range of unstable wave-numbers, but also the magnetic fluctuating energy density reached at the saturation. In effect, the relaxation of anisotropic temperature becomes also more efficient, i.e., faster in time and closer to isotropy.
\end{abstract}


\maketitle

\section{Introduction}\label{Sec.1}
%

Measured in-situ, ion populations in space plasmas are in general found in non-equilibrium, their velocity distributions showing diverse departures from standard Maxwellian \citep{Marsch2006}. The observed non-thermal (or non-equilibrium) distributions may include temperature ($T$) anisotropies, i.e., $T_\perp \neq T_\parallel$ (with $\perp$ and $\parallel$ denoting directions with respect to the {background} magnetic field), beams, and/or suprathermal populations. Ion suprathermal populations are ubiquitous in space plasmas, being often reported by space missions, e.g., in the solar wind \citep{Christon1989, Collier1996}, or in connection with more energetic events, coronal mass ejections (CMEs) \citep{Bamert2004, Tylka2006}, co-rotating interaction regions (CIRs) \citep{Desai1999, Ebert2012, Yu2017}, stream interaction regions (SIRs) \citep{Ebert2012, Yu2018}, and interplanetary shocks \citep{Mazur1992, Lario2019}. In shocks suprathermal ion populations may be a result of the first-order Fermi acceleration \citep{Drury1983}.

In collision-poor space plasmas deviations from isotropy are sources of free energy able to trigger various instabilities. Temperature anisotropy in perpendicular direction, i.e.,  $T_\perp>T_\parallel$, can induce the  electromagnetic ion cyclotron (EMIC) instability \citep{Gary1993, Shaaban2016,Shaaban2016Interplay, Shaaban2017}, which develops as a left-handed (LH) circularly polarized mode, with a maximum growth rate in propagation direction parallel to the ambient magnetic field, i.e., $\bm{k}\times\bm{B}_0=0$. With wave frequency below the proton gyrofrequency $\omega_r<\Omega_p$ \citep{Kennel1966} EMIC wave fluctuations are frequently reported by the observations from, e.g., solar wind \citep{Jian2009}, planetary magnetosphere \citep{Nguyen2007}, Earth's magnetosheath \citep{Anderson1991}, plasma-pause \citep{Fraser1996} and bow shock \citep{Engebretson2013}. In these environments the EMIC fluctuations may also play multiple roles, mainly involving their (resonant) interactions with protons and ions and constraining non-thermal deviations from isotropy of ions. The observations often associate temperature anisotropies of protons with the enhanced low-frequency fluctuations, suggesting a direct cause-effect relationship \citep{Bale2009, Gary2016, Jian2016, Wicks2016}. 

{EMIC instability has  been  extensively investigated considering linear and quasi-linear (QL) approaches of (bi-)Maxwellian populations, e.g., \cite{Gary1993} and \cite{Yoon2012}, with relevance only for the quasi-thermal core of the observed ion velocity distributions (VDs). Suprathermal populations enhance the high-energy tails of the observed distributions, which can markedly deviate from the Maxwellian core and are well described by the Kappa distribution functions \citep{Vasyliunas1968, Christon1989, Collier1996, Pierrard2010}. Recent studies have pointed on the importance of this contrast between the observed Kappa distributions and their Maxwellian core  of lower temperature, in order to outline the presence of suprathermals, and, implicitly, their contribution to kinetic instabilities \citep{Lazar2015Destabilizing, LazarAA2016}. Suggested already in the pioneering work of \cite{Vasyliunas1968}, such a contrasting analysis has not been exploited adequately by the later investigations, which used to compare the results derived for (bi-)Kappa plasmas with those obtained for a (bi-)Maxwellian limit of same temperature\footnote{For extended explanations see \cite{Lazar2015Destabilizing, LazarAA2016}}, preventing a proper description of suprathermals and their effects on the EMIC instability \citep{Lazar2012EMIC, Lazar2017PoP}}. 

{Motivated by these premises, here we refine the QL analysis of EMIC instability in the solar wind conditions, and adopt a proper interpretation of Kappa distributed protons by contrasting to their Maxwellian core, that enables us to describe the effects of suprathermal protons on the extended time evolution of this instability.  Linear studies show a systematic enhancement of the EMIC growth rates in the presence of suprathermals \citep{Shaaban2016}. Moreover, the same realistic interpretation applied to (bi-)Kappa electrons has unveiled a similar systematic stimulation of whistler instability in the presence of suprathermal electrons, with higher growth rates, faster initiation of instability  and higher levels of saturation \citep{ShaabanMNRAS2018, Lazar2019W}.}

In the present paper the anisotropic protons are assumed bi-Kappa distributed, and the effects of suprathermals are outlined from a direct comparison with the Maxwellian limit approaching the thermal core of the distribution (in the absence of suprathermals). QL approach enables us to investigate not only the saturation of growing fluctuations but, also, their back reaction to the velocity distributions. In Section~\ref{Sec.2} we first introduce the anisotropic bi-Kappa model for protons, while the electrons are assumed Maxwellian and initially isotropic; and then provide linear and QL equations used to describe the unstable EMIC solutions. Numerical solutions are derived and discussed for several representative cases in Section \ref{Sec.3}, analyzing the effects of suprathermal protons. The results of the present work are summarized in Section \ref{Sec.4}.

\section{Linear and quasi-linear approaches}\label{Sec.2}
%
We consider a homogeneous and collisionless plasma of anisotropic  suprathermal protons (subscript $p$) and isotropic electrons (subscript $e$). The anisotropic protons are described by the bi-Kappa distribution function \citep{Summers1991}
\begin{align}
 \label{e1}
f_{\kappa,p}\left( v_{\parallel },v_{\perp }\right)=&\frac{1}{\pi ^{3/2} \theta_{\perp p}^{2}~ \theta_{\parallel p}}
\frac{\Gamma\left( \kappa +1\right)}{\kappa^{3/2}\Gamma \left( \kappa -1/2\right)}\nonumber\\
&\times \left[ 1+\frac{v_{\parallel }^{2}}{\kappa~\theta_{ \parallel p }^{2}}
+\frac{v_{\perp }^{2}}{\kappa~\theta_{ \perp p }^{2}}\right] ^{-\kappa-1},
\end{align}
where $\int d^3v f_{\kappa, p}=1$, and $\theta_{\parallel,\perp p}\equiv \theta_{\parallel,\perp p} (t)$ are nominal thermal speeds defined by the temperature components 
\begin{subequations}
\begin{align}\label{e2}
T_{\parallel p}^{\kappa}=&\frac{m_p}{k_B}\int d{\bm v}v_{\parallel}^2f_{\kappa,p}(v_\parallel,v_\perp) =\frac{2\kappa}{2\kappa-3}\frac{m_p \theta^2_{\parallel p}}{2 k_B}\\
T_{\perp p }^{\kappa}=&\frac{m_p}{2k_B}\int d{\bm v}v_{\perp}^2f_{\kappa,p}(v_\parallel,v_\perp) =\frac{2\kappa}{2\kappa-3}\frac{m_p \theta^2_{\perp p}}{2 k_B}
\end{align}
\end{subequations}
Without suprathermals in the high-energy tails, only the bi-Maxwellian core remains, described in the limit $\kappa \rightarrow~\infty$ of the distribution \eqref{e1} given by 
\begin{align}\label{e3}
f_{M,p}\left( v_{\parallel },v_{\perp }\right) =&\frac{1}{\pi
^{3/2}\theta_{\perp p}^{2} ~ \theta_{\parallel p}}\exp \left(
-\frac{v_{\parallel }^{2}} {\theta_{\parallel p}^{2}}-\frac{v_{\perp
}^{2}}{\theta_{\perp p}^{2}}\right),   
\end{align}
with parameters $\theta_{\perp, \parallel p}(t)=\sqrt{2k_B T_{\parallel,\perp p}(t)/m_p}$ (evolving in time $t$ in the quasilinear approach) defined by the temperature components that reduce to 
$\lim_{\kappa \rightarrow \infty}T_{\parallel,\perp p}^{\kappa}=~T_{\parallel,\perp p}$.
{Suprathermals contribute to the kinetic energy budget enhancing the kinetic temperature by comparison to that of  bi-Maxwellian  core \citep{Lazar2015Destabilizing, LazarAA2016}, such that}
\begin{align}
T_{\parallel, \perp p}^{\kappa}=\frac{\kappa}{\kappa-3/2} \;T_{\parallel,\perp p}
> T_{\parallel,\perp p},\label{ne4}
\end{align}  
{and, implicitly, for the proton plasma beta parameters}
\begin{align}
\beta_{\parallel, \perp p}^{\kappa}=\frac{\kappa}{\kappa-3/2} \;\beta_{\parallel,\perp p} 
> \beta_{\parallel,\perp p}. \label{ne5}
\end{align}  
{This realistic interpretation differs from the previous assumptions, e.g., $T_{\parallel, \perp p}^{\kappa}=T_{\parallel, \perp p}$ and $\beta_{\parallel, \perp p}^{\kappa}=\beta_{\parallel, \perp p}$, which minimize the effects of suprathermal populations and lead to questionable results, see, for instance, the irregular variation of the growth rates with $\kappa$ in \cite{Lazar2012EMIC}; and \cite{Lazar2017PoP}.} 

In order to isolate the effects of the suprathermal protons the electrons (subscript $e$) are assumed Maxwellian distributed and initially isotropic
\begin{align}
f_{M,e}(v)=&\frac{1}{\pi
^{3/2} \alpha_{e}^3}\exp \left(-\frac{v_{
}^{2}}{\alpha_{e}^{2}}\right). \nonumber  
\end{align}

The linear dispersion relation of the electromagnetic modes propagating in directions parallel to the background magnetic field (${\bm B}_0$) from non-relativistic dispersion theory reads \citep{Gary1993, Schlickeiser2013}
\begin{align}\label{e4}
 {c^2 k^2\over \omega^2} = 1  +
\sum_{a=p,e} {\omega_{p a}^2 \over \omega^2} \int d\bm{v} {v_\perp\over 2 
\left(\omega - k v_{\parallel} \mp \Omega_a\right)} \nonumber&\\
\times\left((\omega - k v_{\parallel}) {\partial f_{a} \over \partial v_{\perp}} + s
k v_{\perp} {\partial f_{a} \over \partial v_{\parallel}} \right),
\end{align}
where $c$ is the speed of light, $k$ is the wave number, $\omega$ is the wave frequency, $\omega_{p a}= \sqrt{4\pi n_a e^2/m_a}$ and $\Omega_a=~e B_0/m_a c$ are, respectively, the non-relativistic plasma frequency 
and the gyro-frequency of species $a$, $f_a$ the VDF of the species $a$, and $\mp$ denote, respectively, the circular left-handed (LH) or right-handed (RH) polarization. 
For the LH EMIC unstable modes the instantaneous (i.e., calculated with the initial distribution functions \eqref{e1} and \eqref{e3}) dispersion relation (\ref{e4}) reduces to \citep{Shaaban2017}
\begin{align} \label{e5}
\tilde{k}^2&=\mu \left[A_e-1+\left(\frac{A_e~\tilde{\omega} +\left(A_e-1\right)\mu}{\tilde{k} \sqrt{\mu~\beta_e}}\right) Z_{M, e}\left(\frac{\tilde{\omega}+\mu}{\tilde{k} \sqrt{\mu\beta_e}}\right)\right]\nonumber\\
&+A_p-1+\left(\frac{A_p~\tilde{\omega} -\left(A_p-1\right)}{\tilde{k} \sqrt{\beta_p}}\right)Z_{\kappa,p}\left(\frac{\tilde{\omega}-1}{\tilde{k} \sqrt{\beta_p}}\right)
\end{align} 
in terms of normalized wave-number $\tilde{k}=ck/\omega_{p p}$  and frequency $\tilde{\omega}=~\omega/\Omega_p$, $\omega=\omega_r+i\gamma$, proton to electron mass ratio $\mu=m_p/m_e$, temperature anisotropy $A_a=~T_{\perp a}^{\kappa}/T_{\parallel a}^{\kappa}=T_{\perp a}/T_{\parallel a}$, and plasma beta parameter $\beta_{\parallel,\perp a}=~8\pi n_a k_B T_{\parallel \perp  a}/B_0^2$,
\begin{equation}  \label{e6}
Z_{M,e}\left( \xi _{e}^{+}\right) =\frac{1}{\sqrt{\pi}}\int_{-\infty
}^{\infty }\frac{\exp \left(-x^{2}\right) }{x-\xi _{e}^{+}}dx,\
\ \Im \left( \xi _{e}^{+}\right) >0, 
\end{equation}
{is the standard plasma dispersion function \citep{Fried1961} of argument $\xi_e^{+}=(\omega-|\Omega_e|)/(k \alpha_{\parallel e})$, and} 
\begin{align} \label{e7}
 Z_{\kappa,p}\left( \xi_{p}^{-}\right) =&\frac{1}{\pi ^{1/2}\kappa^{1/2}}\frac{\Gamma \left( \kappa \right) }{\Gamma \left(\kappa -1/2\right) }\nonumber\\
     &\int_{-\infty }^{\infty }\frac{\left(1+x^{2}/\kappa \right) ^{-\kappa}}{x-\xi_{p}^{-}}dx,\  \Im \left(\xi _{p}^{- }\right) >0.
\end{align}
{is the modified dispersion function for Kappa-distributed populations \citep{Lazar2008} of argument $\xi_p^{-}=~(\omega+~\Omega_p)/(k \theta_{\parallel p})$ with $\Im \left(\xi _{p}^{-}\right)\equiv \gamma$.}

In the quasi-linear formalism, time evolution of the particle velocity distributions are characterized by the general kinetic equation in the diffusion approximation \citep{Yoon2017R}
\begin{align} \label{e8}
\frac{\partial f_a}{\partial t}&=\frac{i e^2}{4m_a^2 c^2~ v_\perp}\int_{-\infty}^{\infty} 
\frac{dk}{k}\left[ \left(\omega^\ast-k v_\parallel\right)\frac{\partial}{\partial v_\perp}+ 
k v_\perp\frac{\partial}{\partial v_\parallel}\right]\nonumber\\
&\times~\frac{ v_\perp \delta B^2(k, \omega)}{\omega-kv_\parallel-\Omega_a}\left[ 
\left(\omega-k v_\parallel\right)\frac{\partial f_a}{\partial v_\perp}+ k v_\perp
\frac{\partial f_a}{\partial v_\parallel}\right], 
\end{align}
where the wave energy of the fluctuations $\delta B^2(k)$ is described by the wave kinetic equation 
\begin{equation} \label{e9}
\frac{\partial~\delta B^2(k)}{\partial t}=2 \gamma_k \delta B^2(k),
\end{equation}
with the instantaneous growth rate $\gamma_k$ of the EMIC instability derived from the linear dispersion relation \eqref{e5}. 

Dynamical kinetic equations for the perpendicular and parallel velocity moments for protons, and electrons are obtained from \eqref{e8} as follows 
\begin{subequations}\label{e10}
\begin{align}
\frac{dT_{\perp p}}{dt}&=-\frac{e^2}{2m_p c^2}
\int_{-\infty}^{\infty}\frac{dk}{k^2}\langle~ \delta B^2(k)~\rangle\nonumber\\
&\times\left\lbrace\left(2 A_p-1\right)\gamma_k+\text{Im} \frac{2i\gamma-\Omega_p}{k\alpha_{\parallel p}}~ \eta_p^{-}\right\rbrace,\\
\frac{dT_{\parallel p}}{dt}&=\frac{e^2}{m_p c^2}
\int_{-\infty}^{\infty}\frac{dk}{k^2}\langle~ \delta B^2(k)~\rangle\nonumber\\
&\times \left\lbrace A_p~\gamma_k+\text{Im} \frac{\omega-\Omega_p}{k\alpha_{\parallel p}}~ \eta_p^{-}\right\rbrace,\\
\frac{dT_{\perp e}}{dt}&=-\frac{e^2}{2m_e c^2}
\int_{-\infty}^{\infty}\frac{dk}{k^2}\langle~ \delta B^2(k)~\rangle\nonumber\\
&\times\left\lbrace\left(2 A_e-1\right)\gamma_k+\text{Im} \frac{2i\gamma+\Omega_e}{k\alpha_{\parallel e}}~ \eta_e^{+}\right\rbrace,\\
\frac{dT_{\parallel e}}{dt}&=\frac{e^2}{m_e c^2}
\int_{-\infty}^{\infty}\frac{dk}{k^2}\langle~ \delta B^2(k)~\rangle \nonumber\\
&\times\left\lbrace A_e~\gamma_k+\text{Im} \frac{\omega+\Omega_e}{k\alpha_{\parallel}}~ \eta_e^{+}\right\rbrace,
\end{align}
\end{subequations}
with
\begin{align*}
\eta_p^{-}=\left[ A_p~\omega-\Omega_p\left(A_p-1\right)\right]Z_{\kappa,p}\left(\xi_p^{-}\right),\\
\eta_p^{+}=\left[ A_e~\omega+\Omega_e\left(A_e-1\right)\right]Z_{M,e}\left(\xi_e^{+}\right).
\end{align*}

Using the same normalized quantities from Eq.~\eqref{e5}, the dynamical kinetic equations \eqref{e10} can be rewritten as
\begin{subequations}\label{e11}
\begin{align}
\frac{d\beta_{\perp p}}{d\tau}=&-\int\frac{d\tilde{k}}{\tilde{k}^2} W(\tilde{k})\nonumber\\
&\times\left\lbrace\left(2 A_p-1\right)\tilde{\gamma}+\text{Im} \frac{2i\tilde{\gamma}-1}{\tilde{k}\sqrt{\beta_{\parallel p}}}~\tilde{\eta}_p^{-}\right\rbrace,\\
\frac{d\beta_{\parallel p}}{d\tau}=&2\int\frac{d\tilde{k}}{\tilde{k}^2} W(\tilde{k})\left\lbrace  A_p~\tilde{\gamma}+\text{Im} \frac{\tilde{\omega}-1}{\tilde{k}\sqrt{\beta_{\parallel p}}}~\tilde{\eta}_p^{-}\right\rbrace,\\
\frac{d\beta_{\perp e}}{d\tau}=&-\int\frac{d\tilde{k}}{\tilde{k}^2} W(\tilde{k})\nonumber\\
&\times\left\lbrace \mu \left(2 A_e-1\right)\tilde{\gamma}+\text{Im} \frac{2i\tilde{\gamma}+\mu}{\tilde{k}\sqrt{\beta_{\parallel c}}}~\tilde{\eta}_e^{+}\right\rbrace,\\
\frac{d\beta_{\parallel e}}{d\tau}=&2\int\frac{d\tilde{k}}{\tilde{k}^2} W(\tilde{k})\left\lbrace \mu~A_e~\tilde{\gamma}+\text{Im} \frac{\tilde{\omega}+\mu}{\tilde{k}\sqrt{\beta_{\parallel e}}}~\tilde{\eta}_e^{+}\right\rbrace
\end{align}
\end{subequations}
with
\begin{align*}
\tilde{\eta}_p^{-}=&\left[A_p~\tilde{\omega}-\left(A_e-1\right)\right]Z_{p,\kappa}\left(\xi_p^{-} \right),\\
\tilde{\eta}_e^{+}=&\sqrt{\mu}\left[ A_e~\tilde{\omega}+\left(A_e-1\right)\mu\right]Z_{e}\left(\xi_e^+\right),
\end{align*}
and 
\begin{align}\label{e12}
\frac{\partial~W(\tilde{k})}{\partial \tau}=2~\tilde{\gamma}~ W(\tilde{k}).
\end{align}
where $W(\tilde{k})=\delta B^2(\tilde{k})/B_0^2$ is the normalized spectral wave energy density, and $\tau=~\Omega_p~t$.
%
%
%
\begin{figure}[ht]
\centering 
\includegraphics[width=0.4\textwidth, trim=2.4cm 1.3cm 2.cm 2.6cm, clip]{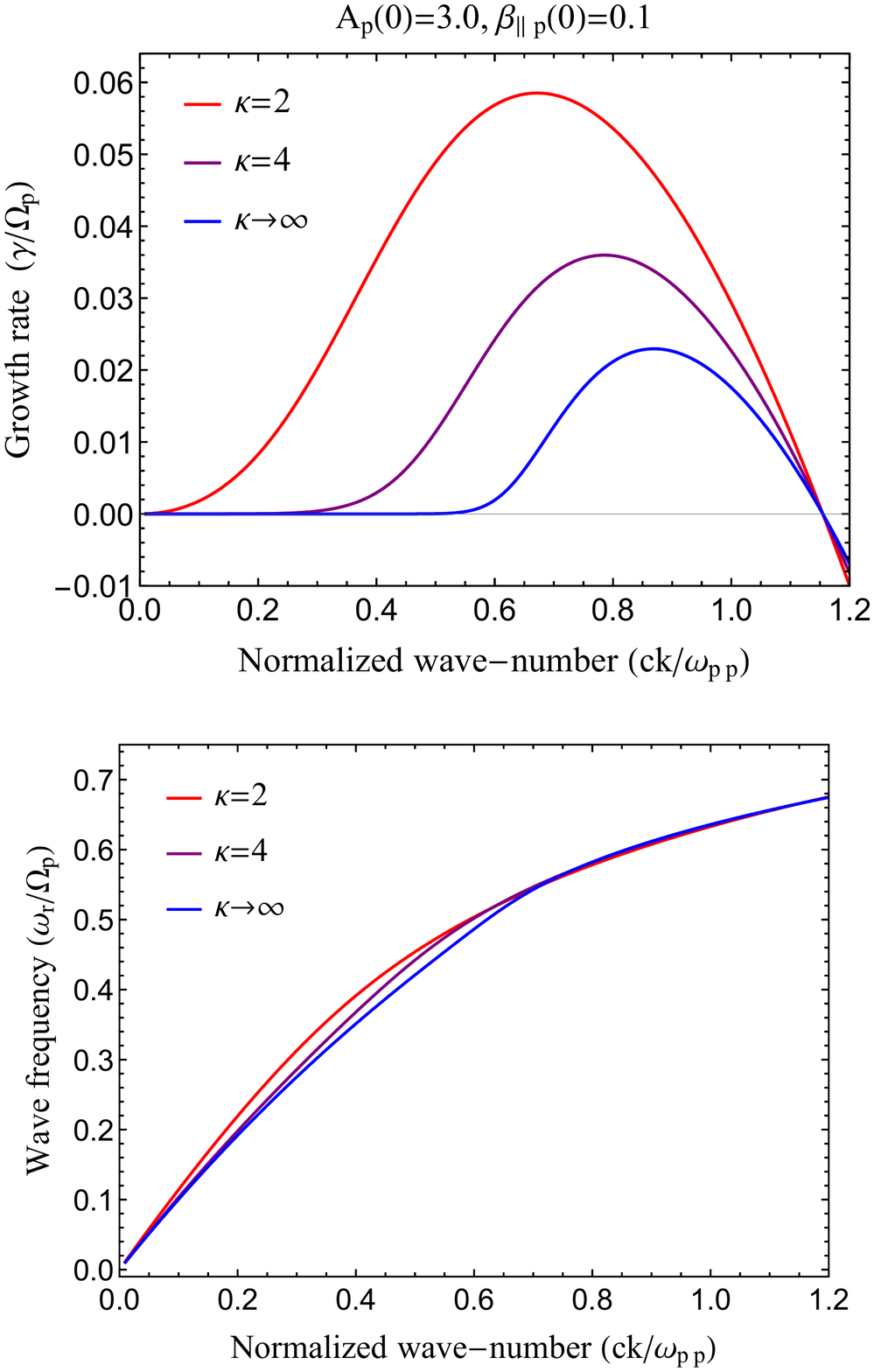}
\caption{{Effects of the proton power-index $\kappa$ on the growth rates (top panel) and wave frequencies (bottom panel) of EMIC instability for plasma parameters given in case 1.}}
\label{f1}
\end{figure}

\section{Numerical stability analysis}\label{Sec.3}

In this section, we solve numerically the QL equations \eqref{e11} and \eqref{e12} for three sets of initial plasma parameters (i.e., at $\tau=0$)
\begin{itemize}
\item Case 1: $A_p(0)=3.0$, $\beta_{\parallel p}(0)=0.1$, and $\kappa=2, 4, \infty$
\item Case 2: $A_p(0)=3.0$, $\beta_{\parallel p}(0)=0.5$, and $\kappa=2, \infty$
\item Case 3: $A_p(0)=3.0$, $\beta_{\parallel p}(0)=2.0$, and $\kappa=2,\infty$,
\end{itemize}
corresponding to small (case 1), intermediate (case 2), and large (case 3) parallel plasma betas. Other common parameters $A_e(0)=1.0$, $\beta_{\parallel e}(0)=\beta_{\parallel e}(0)$, initial spectral wave energy density of level $W(\tilde{k},0)=10^{-6}$, and proton Alfv{\'e}n speed $v_A =~ 2\times 10^{-4} c$.
{The initial macroscopic plasma parameters are selected to cover conditions relevant for the solar wind space plasmas, and outline the effects of suprathermal protons, by contrasting the unstable solutions derived for the observed bi-Kappa distribution ($\kappa=2$) with those obtained for a bi-Maxwellian core ($\kappa\rightarrow \infty$). Other combinations of initial plasma parameters can also be used, as we discuss below in Figure~\ref{f7}.}

\begin{figure}[t]
\centering 
\includegraphics[width=0.4\textwidth, trim=5.2cm 5.cm 5.2cm 2.6cm, clip]{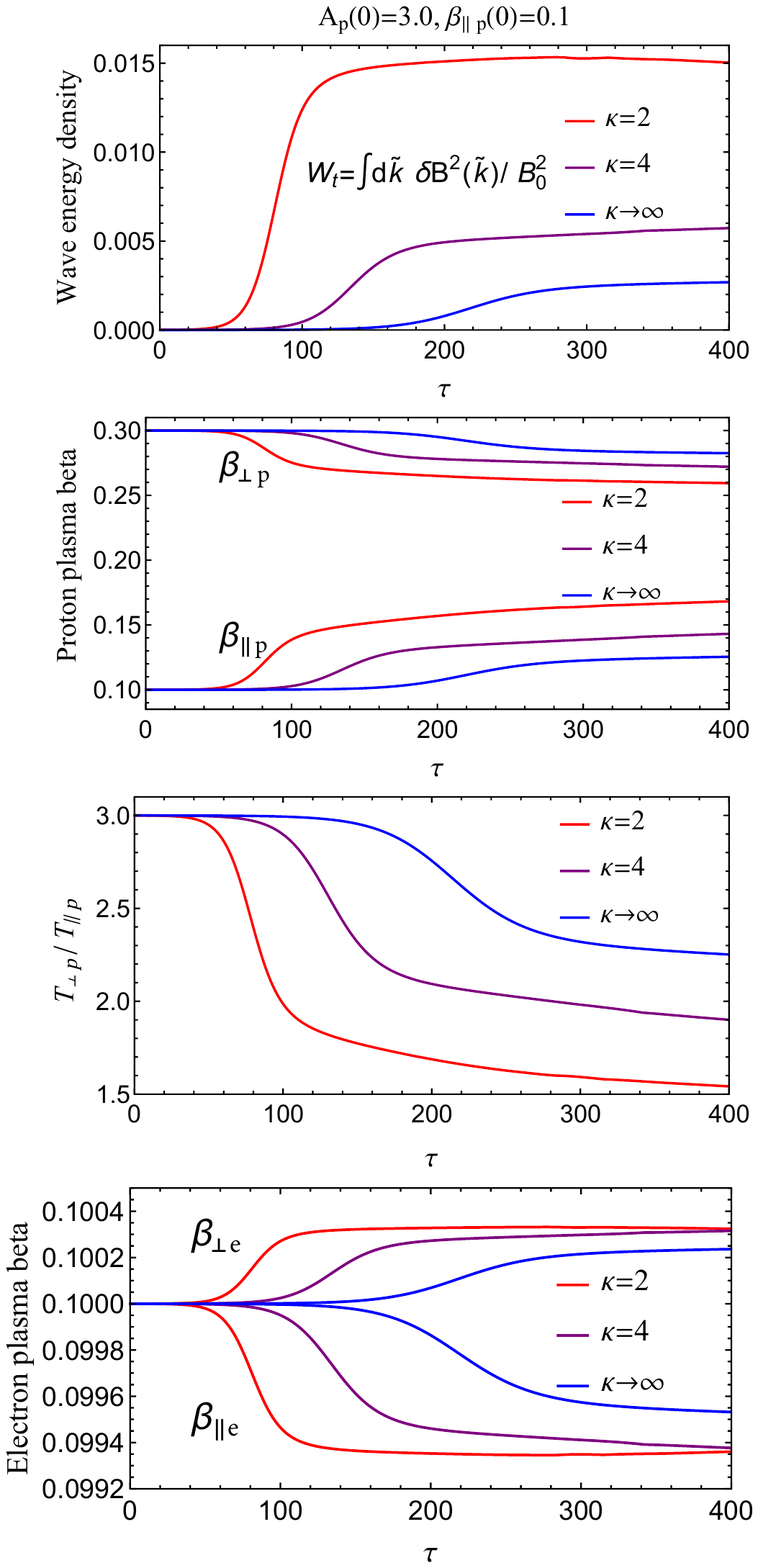}
\caption{Effects of the proton power-index $\kappa=2,4$ and $\infty$ on magnetic wave energy density $W_t$ (top), plasma betas $\beta_{\perp,\parallel p}$ (middle-first), the relaxation of proton temperature anisotropy $A_p$ (middle-second) and electron plasma betas $\beta_{\perp,\parallel e}$ (bottom) for the same plasma parameters in Figure~\ref{f1}.}
\label{f2}
\end{figure}

\begin{figure*}[ht]
\centering 
\includegraphics[width=0.9\textwidth, trim=2.8cm 15.5cm 2.2cm 2.9cm, clip]{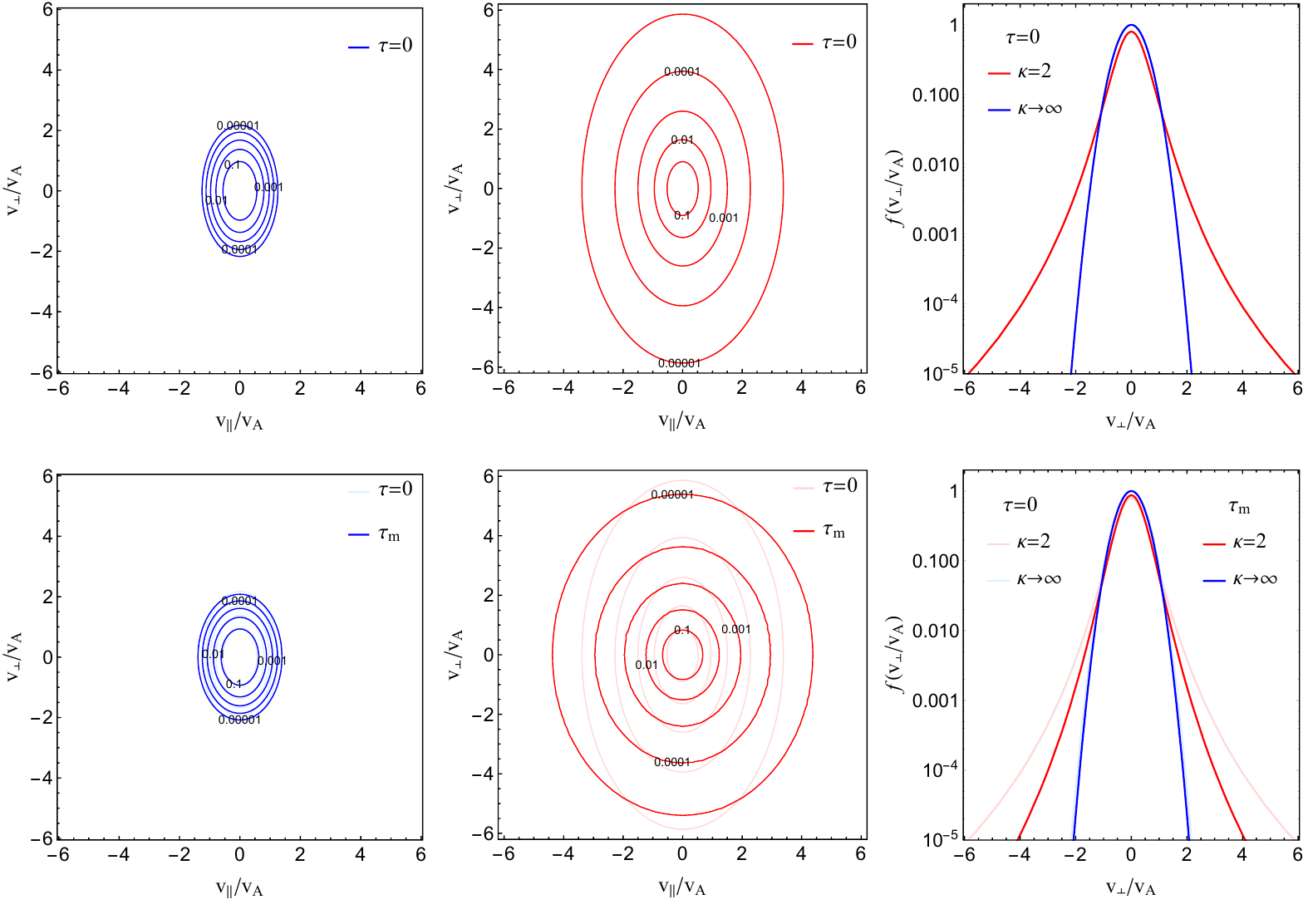}
\caption{Contours of proton velocity distribution, for the Maxwellian core (left panel, blue) and total Kappa (middle, red, $\kappa=2$), and the corresponding perpendicular cuts (right), initially at $\tau=0$ (top) and after saturation $\tau=\tau_m$ (bottom).}
\label{f3}
\end{figure*}

From a guiding linear analysis here we explicitly show only the results for case 1, deriving from \eqref{e5} the wave-number dispersion curves for the growth rates and wave frequency. These (unstable) EMIC solutions are presented in Figure \ref{f1}, for two distinct cases with $\kappa=$2, and 4, and Maxwellian limit $\kappa \to\infty$. Growth rates (top panel) are markedly  enhanced by increasing the presence of suprathermal protons, i.e., lowering $\kappa$. The maximum (peaking) growth rate is approximately $1.5$ times higher for $\kappa=4$ (purple), and roughly 3 times higher for $\kappa=2$ (red) than that for bi-Maxwellian protons (blue). {Note that peaks of the growth rates shift to lower wavenumbers by increasing the abundance of suprathermal protons. If thermal velocity of the resonant protons move to higher energies, the resonant modes (satisfying $|\xi_p^{-}|\sim 1$), and implicitly the maximum growth rates ($\gamma_m$) move to lower wavenumbers, accordingly to 
\begin{align}
    |\xi_p^{-1}|\equiv\left|\frac{\tilde{\omega}_r+i \tilde{\gamma}_m-~1}{\tilde{k}_m v_{res}/v_A}\right| \sim 1.
\end{align}
where $\tilde{\gamma}_m\equiv \gamma_m/\Omega_p$ is the value of the growth rate peak, $\tilde{k}_m\equiv ck_m/\omega_{p p}$, and $\tilde{\omega}_r\equiv \omega_r/\Omega_p$ are the corresponding wave-number and wave frequency, respectively.}
{Wave frequencies of EMIC instability (bottom panel) show minor variations with decreasing the power exponent $\kappa$.}

Beyond these effects on linear properties of EMIC instability, here we  describe the QL evolution for both the fluctuating energy in \eqref{e12}, and the anisotropy of protons in \eqref{e11}. 
Figure~\ref{f2} presents the QL solutions for the initial plasma parameters of case 1, and for $\kappa=2$ (red), $4$ (purple), and $\kappa \to \infty$ (blue). {It displays temporal profiles for the magnetic wave-energy density $W_t=\int d\tilde{k}~\delta B_0^2(\tilde{k})/B_0^2$ (top panel), perpendicular ($\perp$) and parallel ($\parallel$) plasma betas for protons $\beta_{\perp, \parallel p}$ (middle-first), the proton temperature anisotropy $A_p$ (middle-second), and plasma betas for electrons $\beta_{\perp, \parallel e}$ (bottom)}. Time is normalized as $\tau=~\Omega_p t$. These results confirm predictions from linear theory, and additionally show a long-run stimulation of instability in the presence of suprathermals,  and the effects of growing fluctuations back on protons. One may first notice an enhancement of the resulting magnetic wave-energy density ($W_t$) of EMIC fluctuations, with a faster initiation, a steeper growing profile, and a higher level of the saturation. 

\begin{figure}[h]
\centering 
\includegraphics[width=0.4\textwidth]{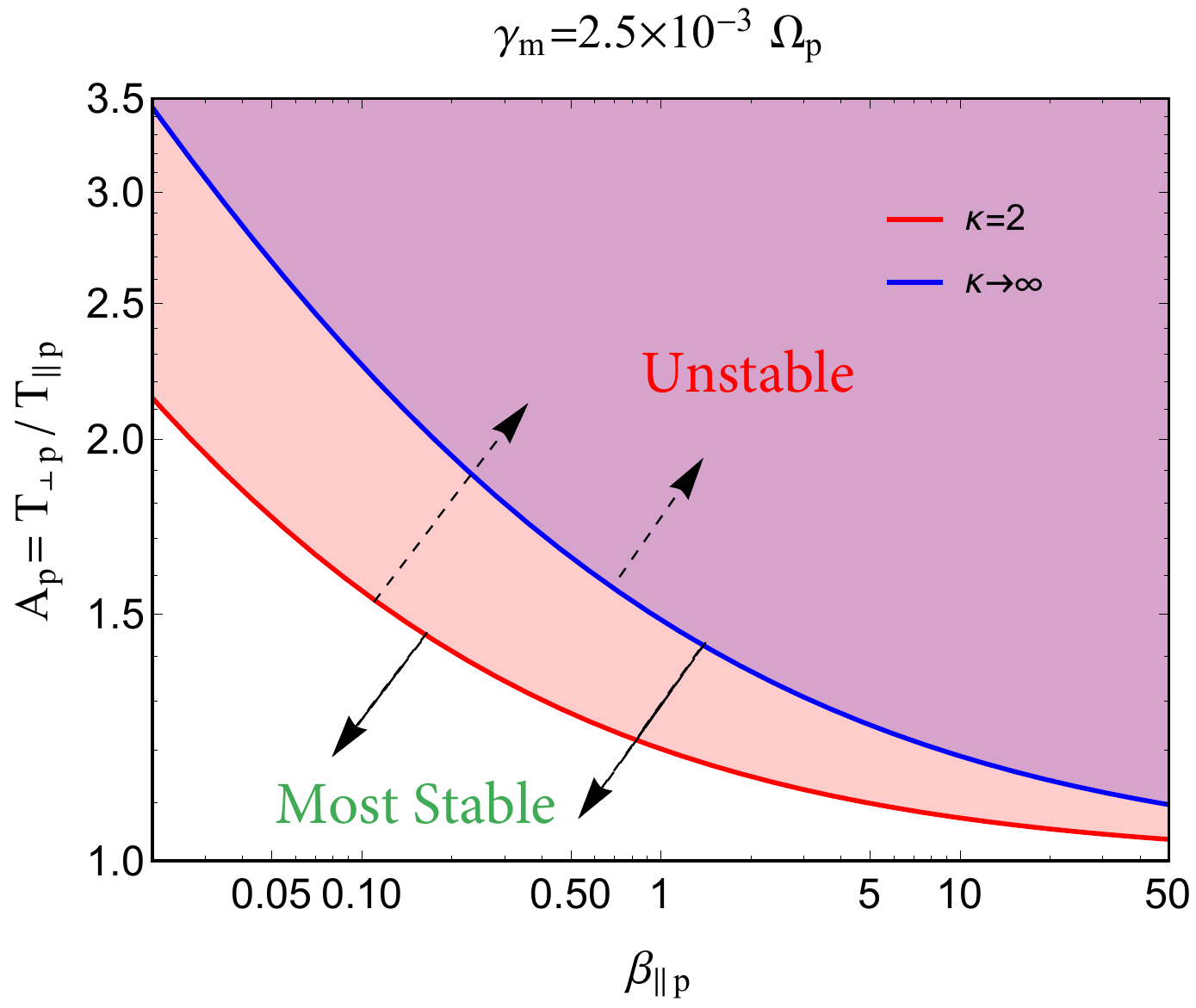}
\caption{Anisotropy thresholds for the EMIC instability (maximum growth rate $\gamma_m/\Omega_p=2.5\times 10^{-3}$) driven by bi-Maxwellian (blue) and bi-Kappa (red) distributed protons.}
\label{f4}
\end{figure}

\begin{figure*}[t]
\centering 
\includegraphics[width=0.9\textwidth, trim=2.6cm 14.2cm 2.2cm 2.5cm, clip]{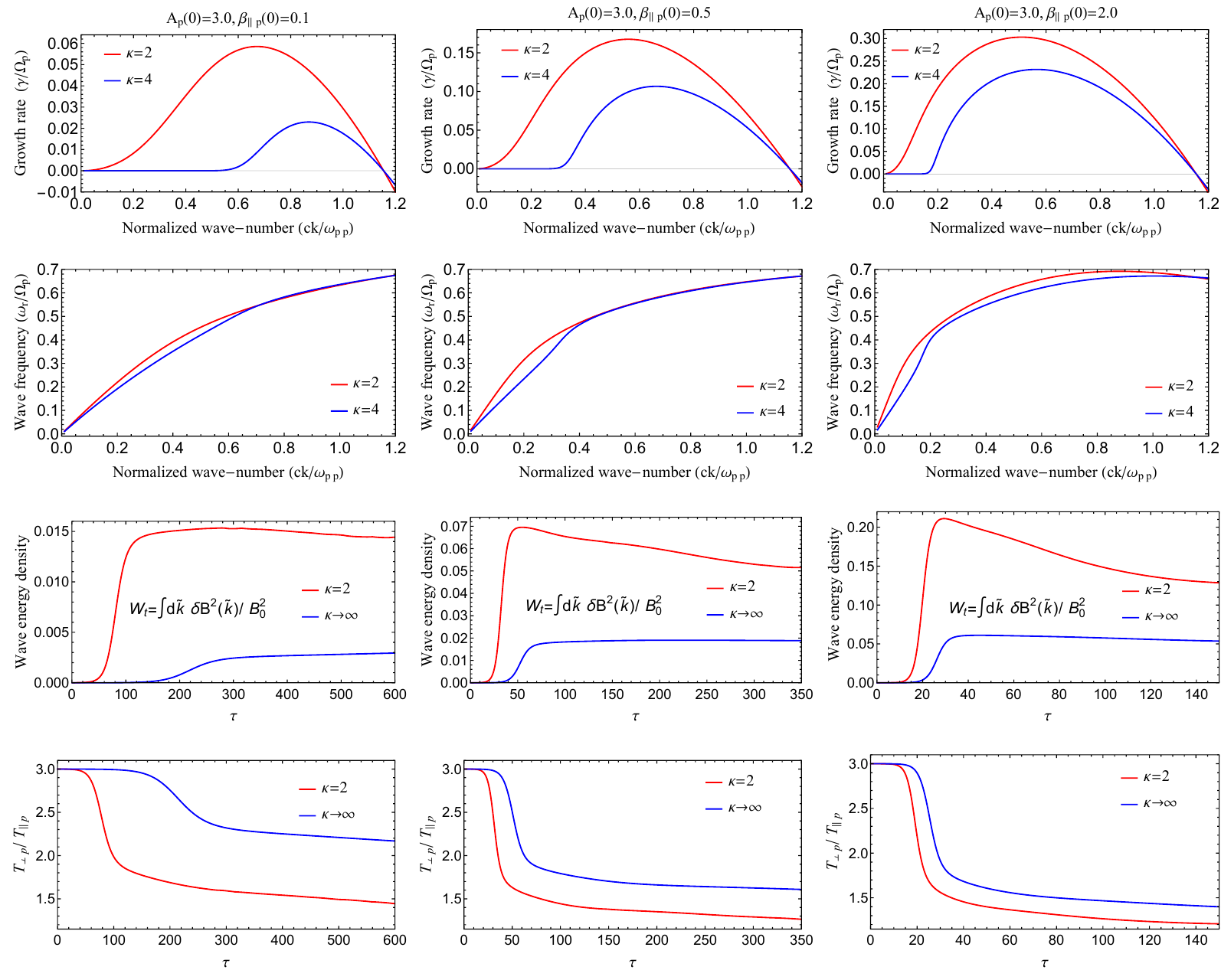}
\caption{Effects of the suprathermal protons on the growth rate $\gamma/\Omega_p$ (top), wave-frequency $\omega_r$ (middle-first), magnetic wave energy density $W_t$ (middle-second) and the relaxation of proton temperature anisotropy $A_p$ (bottom) for different plasma beta $\beta_p=0.1$ (left), 0.5 (middle) and 2.0 (right).}
\label{f5}
\end{figure*}
EMIC fluctuations regulate the initial temperature anisotropy of protons, as reflected by the decrease of $\beta_{\perp p}$ and the increase of $\beta_{\parallel p}$ (middle-first panel), suggesting a perpendicular cooling combined with a parallel heating of protons. After the saturation, i.e., $\tau > \tau_m$, protons are less anisotropic with $A_p(\tau_m)<A_p(0)$ (middle-second panel). 
{Initially isotropic electrons ($A_e(0)=1$) develop a small anisotropy by gaining a modest energy in perpendicular direction, i.e., $A_e(\tau_m)\sim 1.01$ after saturation, possibly from anomalous resonance with the LH EMIC fluctuations, which usually involves only a small component of electron population \citep{Tsurutani1997}}. 
In the presence of suprathermal protons (red lines for $\kappa=2$), the magnetic wave energy reaches a maximum level much higher than that obtained for bi-Maxwellian limit $\kappa \rightarrow \infty$. In  turn, the enhanced fluctuations of EMIC instability lead to more pronounced effects on the proton VDF, i.e., a faster and deeper relaxation approaching the quasi-stable state of anisotropic temperature.

Most prominent are the effects obtained for $\kappa=2$. In Figure~\ref{f3} we display the normalized proton VDF corresponding to $\kappa = 2$ (red), by comparison with that corresponding to $\kappa \rightarrow \infty$ (blue), as contours in ($v_\perp/v_A, v_\parallel/v_A$)--space (left and middle panels) and perpendicular cuts (right panels), for the initial $\tau=0$ (top panels) and final (maximum) time step $\tau=\tau_m$ after saturation (bottom panels). For all snapshots contour levels $10^{-1},10^{-2}, 10^{-3}, 10^{-4}$, and $10^{-5}$ of $f_{max}=1$ are shown. As one can see in the bottom panels, compared to the initial state (see also the light-blue and light-red contours in the background) the proton VDFs {become less anisotropic in perpendicular direction and more stable against EMIC instability}. It is obvious that the final state of the proton VDF for $\kappa=2$ is much less anisotropic than that obtained for $\kappa \rightarrow \infty$, and therefore more stable.   

\begin{figure*}[t]
\centering 
\includegraphics[width=1\textwidth]{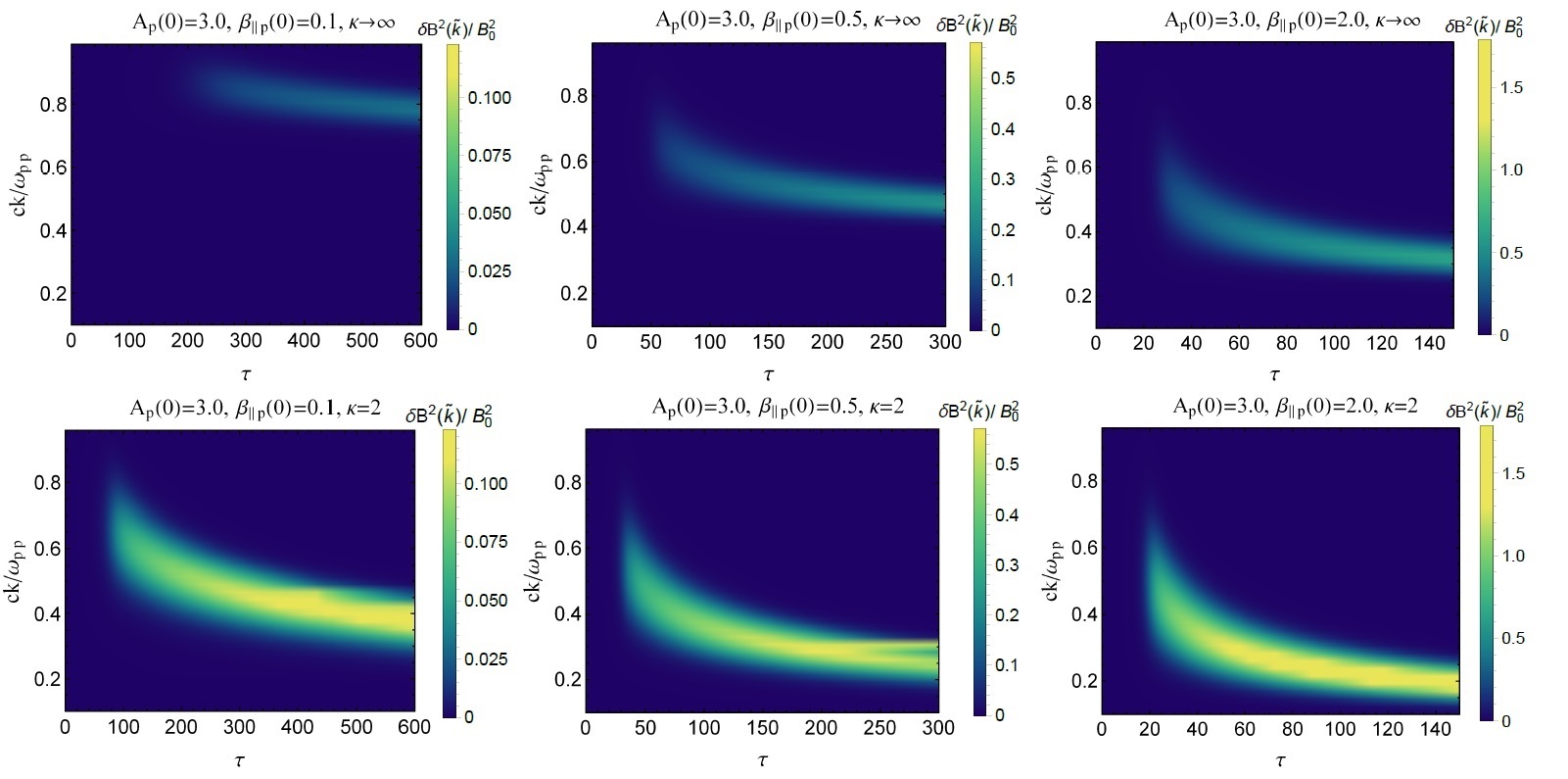}
\caption{Effects of the suprathermal protons on the temporal evolution of the wave-number spectra of the {magnetic wave energy density $W(\tilde{k})=\delta B^2(\tilde{k})/B_0^2$. }The plasma parameters are mentioned in each panel.}
\label{f6}
\end{figure*}

Figure~\ref{f4} shows the anisotropy thresholds derived as a function of $\beta_{p,\parallel}$ for the maximum growth rate $\gamma_m=2.5\times 10^{-3}$, close to the marginal stability of EMIC modes. These thresholds provide a comprehensive picture for the effects of suprathermal protons on the extended unstable regimes of EMIC modes. These thresholds are obtained from linear dispersion relation \eqref{e5} and are fitted by
\begin{align}
A_p=1+\frac{a}{\beta_{\parallel p}^{b}}
\end{align}
with fitting parameters $(a, b)= (0.20,0.44)$ for $\kappa=2$ (red curve) and $(a,b)=(0.48, 0.41)$ for $\kappa \rightarrow \infty$ (blue).
Thresholds decrease with increasing $\beta_\parallel$, extending the unstable regime of EMIC modes to lower anisotropies $A_p\gtrsim 1$. For hotter plasmas ($T_{\parallel p}\propto \beta _{\parallel p}$) EMIC modes need lower anisotropies to become unstable, see e.g., \cite{Shaaban2016, Shaaban2017}. Solid arrows indicate the most stable EMIC regimes below the thresholds, while the unstable EMIC regimes are located above the thresholds, as pointed with dashed arrows. The effects of suprathermal protons are again outlined {by contrasting the anisotropy thresholds} for $\kappa=2$ (red curve) and for a bi-Maxwellian protons (blue). The anisotropy threshold is markedly lowered in the presence of suprathermal protons (i.e. for $\kappa=2$), increasing the unstable regime of EMIC modes (red shaded area) and confirming the instability stimulation. Physical interpretation extracted from Figure~\ref{f4} is obvious: in the presence of suprathermal protons EMIC modes need lower anisotropies to become unstable. 

\begin{figure*}[t]
\centering 
\includegraphics[width=0.48\textwidth]{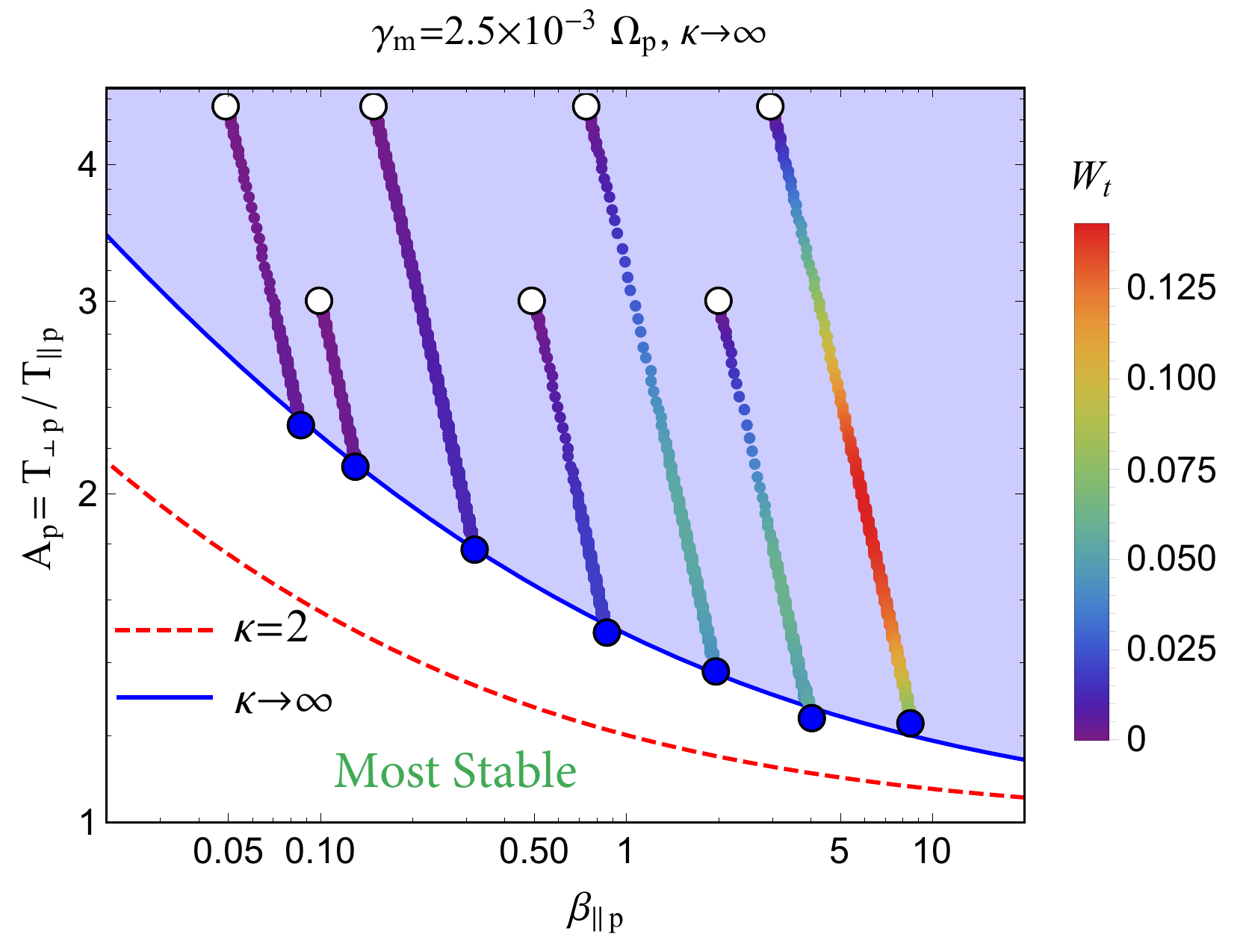}~~~~
\includegraphics[width=0.48\textwidth]{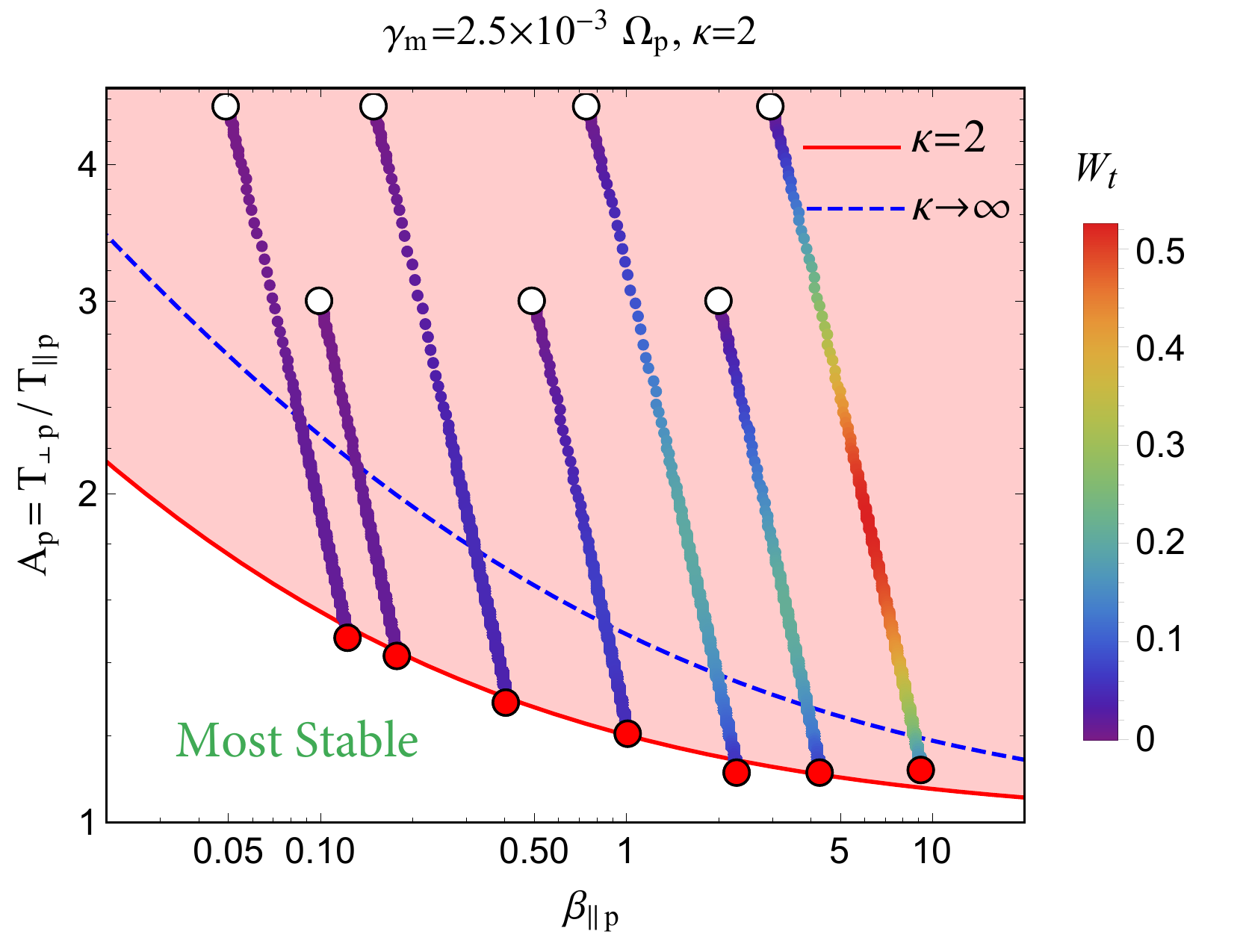}
\caption{Quasi-linear dynamical decreasing paths for the bi-Maxwellian (left) and bi-Kappa (right) distributed protons in ($A_p$, $\beta_{\parallel p}$)--space.}
\label{f7}
\end{figure*}

Starting from the premises that linear growth rates are stimulated by increasing the temperature (plasma beta) of protons, or the presence of their suprathermal component, or both, in Figure~\ref{f5} we describe the influence of these factors on the long-run QL evolution of EMIC fluctuations and the relaxation of proton anisotropy. Three distinct cases are considered, corresponding to different initial  $\beta_{\parallel p}(0)=0.1$ (left), $\beta_{\parallel p}(0)=0.5$ (middle), and $\beta_{\parallel p}(0)=0.2$ (right), known already as cases 1, 2 and 3, respectively. For all cases the growth rates $(\gamma/\Omega_p$) of EMIC instability are systematically enhanced by suprathermal protons, i.e., for $\kappa=2$, (red curves in the top panels). The corresponding wave-frequencies show only minor variations with $\kappa$ (middle-first panels).  Consistent with this stimulation of growth rates, QL solutions show a similar systematic enhancement of the magnetic wave-energy density $W_t$ in the presence of suprathermal protons (middle-second panels), with saturation levels increasing with increasing the initial plasma beta. The enhanced fluctuations resulting from these cumulative effects determine a more pronounced action on anisotropic protons, leading to a faster and deeper relaxation of their anisotropy $T_{\perp p}/T_{\parallel p}(\tau)\gtrsim 1$ (bottom panels). Similar to Figure~\ref{f2} the variations of electron plasma betas remain minor, and are not shown here. 
Furthermore, in Figure~\ref{f6} we show the same temporal evolution for the spectral wave energy density $W(\tilde{k})=~\delta B^2(\tilde{k})/B_0^2$ as a function of wave-number. The intensity of the spectral wave energy $W(\tilde{k})$, codded with colors in the right-hand bars, is stimulated by the abundance of suprathermal protons, i.e., $\kappa=2$, for all wave-numbers (bottom panels). This effect is boosted by increasing $\beta_{\parallel p}$, supporting the results in Figure~\ref{f5}.

Figure~\ref{f7} presents a comparison between theoretical predictions from linear and quasi-linear approaches. For the sake of completeness, in Figure~\ref{f7} we have considered four additional cases of distinct initial parameters, including a different temperature anisotropy and more different parallel plasma betas. {The new cases correspond to an initial anisotropy $A_p(0)=4.5$ and four different $\beta_{\parallel p}(0)=0.05$, $0.15$, $0.75$, $3.0$.} The QL evolution for $A_p$ and $\beta_{\parallel p}$ are displayed as dynamical paths beginning at the initial conditions, indicated by white circles, and stopping after the saturation at the final positions, marked with blue circles for  $\kappa \rightarrow \infty$ and red circles for $\kappa=2$. As expected, for all cases the initial anisotropies are reduced in time toward less unstable states closer to marginal stability, and final states align exactly along the anisotropy thresholds predicted by the linear theory. This relaxation of the proton anisotropy is a direct consequence of the enhanced EMIC fluctuations and their magnetic energy density $W_t$, color-coded with a rainbow color scheme. Note that suprathermal protons, i.e., for $\kappa=2$ (right panel) determine longer dynamical paths for the relaxation of protons, and, implicitly, a deeper relaxation of their anisotropic temperature. 
 
\section{Conclusions}\label{Sec.4}
We have presented a {refined} QL analysis of EMIC instability driven by the anisotropic temperatures of bi-Kappa distributed protons, of high relevance in space plasmas, e.g., solar wind, CMEs, planetary magnetosphere, and shocks. Recent studies have pointed on the importance of a realistic interpretation of Kappa models in order to properly describe the effects of suprathermal populations \citep{Lazar2015Destabilizing, LazarAA2016, Shaaban2016, Shaaban2020}. Linear studies of EMIC instabilities have predicted a systematic enhancement of the (maximum) growth rates or the range of unstable wave-numbers in the presence of the suprathermal protons, and a significant decrease of threshold anisotropy \citep{Shaaban2016}. In the present paper we have described the long-run QL evolution of the growing fluctuations and consequences of their interaction with plasma particles, leading to the relaxation of anisotropic temperature. Within a realistic Kappa interpretation, as adopted in the present work, suprathermal protons contribute with an excess of free energy that systemically stimulates the EMIC instability.

In order to highlight these effects we have contrasted the results obtained for bi-Kappa distribution, in the presence of suprathermals, with those obtained for a bi-Maxwellian approaching the  quasi-thermal core. 
Thus, the results in section \ref{Sec.3} show the effects of suprathermal protons on the time evolution of EMIC fluctuation and macroscopic plasma parameters, including the saturation levels of magnetic energy density, and the relaxation of proton velocity distributions. 
The enhanced fluctuations with higher growth rates predicted by linear theory in the presence of suprathermals (Figure~\ref{f1}) become more robust reaching higher levels of magnetic-wave energy density. The corresponding relaxation of anisotropic protons becomes also faster leading to a deeper relaxation, see Figures~\ref{f2} and \ref{f3}. Previous studies underestimate these effects due to a simplified (bi-)Kappa interpretation, see, e.g., \citep{Lazar2017PoP}. 
Figures~\ref{f1}-\ref{f3} provide a detailed description of the stimulative effects of suprathermal protons on the EMIC instability, in linear and quasi-linear phases, but for a single value of plasma beta ($\beta_{\parallel p}=0.1$). In Figure~\ref{f4}-\ref{f6} we have studied these effects for an extended range of the proton plasma beta, i.e., $0.02\leqslant \beta_{\parallel p}\leqslant 50$. The stimulative effects of suprathermal  protons remain systematic for the full range of these values. The anisotropy threshold in Figure~\ref{f4} is lowered and the unstable regime is considerably enhanced due to the abundance of suprathermal protons (i.e., decreasing $\kappa$), and with increasing $\beta_{\parallel p}$. 

For the QL analysis we have chosen three distinct cases corresponding to low, intermediate and high beta conditions ($\beta_{\parallel p}=0.1, 0.5$, and 2.0). The results show a systematic stimulation of the spectral and total magnetic energy density of EMIC fluctuations, and in turn a faster and more efficient relaxation for the proton distribution in the presence of suprathermals (Figure~\ref{f5} and \ref{f6}). We have performed a comparative analysis between the anisotropy thresholds obtained from linear theory with the quasi-linear dynamical paths of the temperature anisotropies, which are obtained for seven distinct sets of initial parameters in Figure~\ref{f7}. Predictions from linear and quasi-linear approaches are in perfect agreement.    

We conclude stating that suprathermal protons have an important impact, stimulating the growth of EMIC instability in both linear and quasi-linear phases, showing not only a faster developing but also a more efficient relaxation of the anisotropic protons. Comparing to the results for more idealized bi-Maxwellian plasmas which ignore the effects of suprathermal protons, our results show an extended implication/ EMIC instability in the non-equilibrium Kappa distributed plasmas from space, where instabilities are expected to play a major role bounding the non-thermal departures from isotropy.
%
\section*{Acknowledgements}
%
The authors acknowledge support from the Katholieke Universiteit Leuven, Ruhr-University Bochum and Christian-Albrechts-Universit\"at Kiel. These results were obtained in the framework of the projects SCHL 201/35-1 (DFG-German Research Foundation), GOA/2015-014 (KU Leuven), G0A2316N (FWO-Vlaanderen). S.M.Shaaban acknowledges the Alexander-von-Humboldt Research Fellowship and thanks Prof. Robert Wimmer-Schweingruber for his kind hospitality and support.
%
\section*{DATA AVAILABILITY}
%
The data that support the findings of this study are available
from the corresponding author upon  request.
%
\bibliographystyle {agsm}
\bibliography{allpapers}
\end{document}